\documentclass[prl,twocolumn]{revtex4}

\usepackage{graphicx}
\usepackage{dcolumn}
\usepackage{bm}

\newcommand{\bo}{\raise-1mm\hbox{\Large$\Box$}}

\newcommand{\rr}{\mathbf{r}}
\newcommand{\kk}{\mathbf{k}}
\usepackage{epsfig}
\usepackage{latexsym}
\usepackage{mathrsfs}
\usepackage{setspace}
\usepackage{amsmath}
\usepackage{floatflt}
\usepackage{wrapfig}

\begin{document}

\title{Critical superfluid velocity in a trapped dipolar gas}

\author{Ryan M. Wilson$^1$\email{rmw@colorado.edu}}
\author{Shai Ronen$^{2,1}$}
\author{John L. Bohn$^1$}

\affiliation{$^1$JILA and Department of Physics, University of Colorado, Boulder, Colorado 80309-0440, USA}
\affiliation{$^2$University of Innsbruck and Institute for Quantum Optics and Quantum Information, Innsbruck, Austria}

\date{\today}

\begin{abstract}
We investigate the superfluid properties of a dipolar Bose-Einstein condensate (BEC) in a fully three-dimensional trap.  Specifically, we estimate a superfluid critical velocity for this system by applying the Landau criterion to its discrete quasiparticle spectrum.  We test this critical velocity by direct numerical simulation of condensate depletion as a blue-detuned laser moves through the condensate.  In both cases, the presence of the roton in the spectrum serves to lower the critical velocity beyond a critical particle number.  Since the shape of the dispersion, and hence the roton minimum, is tunable as a function of particle number, we thereby propose an experiment that can simultaneously measure the Landau critical velocity of a dipolar BEC and demonstrate the presence of the roton in this system.
\end{abstract}

\maketitle

Liquid $^4\mathrm{He}$  was the first experimentally accessible system to exhibit dissipationless flow at low temperature, i.e., to demonstrate the existence of superfluidity in a quantum system.  Landau famously explained this phenomenon by identifying a critical velocity $v_L$ below which elementary excitations in the fluid could not be excited while conserving energy and momentum \cite{Landau41}.  Because of this connection to the spectrum of elementary excitations, the Landau critical velocity can be expressed in terms of the fluid's dispersion relation $\omega (k)$ as
\begin{equation}
\label{landauvc}
v_L = \min{\left[\frac{\omega (k)}{k}\right]}.
\end{equation}
Remarkably, the Landau critical velocity $v_L$ does not coincide with the speed of sound in liquid helium, but is smaller due to the existence of an anomalously low-energy roton mode at wave vector $k \sim  \AA^{-1}$.  This critical velocity was ultimately verified in experiments of ion drift velocity in liquid $^4$He~\cite{Allum77}.

More recently, a new class of superfluids has been produced in the form of Bose-Einstein condensates (BECs) of ultracold atomic  gases.  These gases have a distinct advantage over liquid helium in that they are dilute and hence easily characterized in terms of microscopic interactions.  In particular, their critical velocity is nominally given by the speed of sound in the center of the gas, which can be easily calculated from the density and the s-wave scattering length of the constituent atoms.   Early experiments at MIT sought to measure $v_L$ in a BEC of sodium atoms by stirring the condensate with a blue-detuned laser~\cite{Raman99,Onofrio00}.  However, these experiments measured a critical velocity for spinning off vortices rather than the true Landau critical velocity.  This is  a generic feature of such experiments in which the size of the object (in this case, the blue-detuned laser) is large compared to the healing length of the gas~\cite{Frisch92,Winiecki00,Stiessberger00,Jackson00}.

Still more recently, atomic BECs have been created whose constituent atoms possess magnetic dipole moments large enough to influence the condensate~\cite{Griesmaier05a,Werner05}.  These gases present a middle ground between atomic BECs and dense superfluid helium.  Namely, the dipolar BEC (DBEC) is dilute enough to be understood in detail, yet its spectrum may exhibit roton features in prolate traps, like those of liquid He~\cite{Santos03a}.  The characteristic momentum of such a roton is set by the geometry of the trap in which it is held, whereas its energy is controlled by the density of dipoles, as well as the magnitude of the dipole moment~\cite{LahayeRev09}.  Thus, by Eq.~(\ref{landauvc}), the Landau critical velocity is completely under the control of the experimentalist.  In contrast, $v_L$ in $^4$He can be only weakly modified by changing the pressure of the liquid~\cite{Dietrich72}.  Thus, the DBEC provides an unprecedented opportunity to explore the fundamental relationship between the roton dispersion and superfluidity.

In this Letter we model an experiment on a DBEC similar to the MIT experiments.  We consider a blue-detuned laser sweeping through a DBEC at a constant velocity, then compute the resulting condensate depletion due to the excitation of quasiparticles.  We find an onset of depletion at a critical velocity that is near the Landau critical velocity at low densities.  At higher densities, where the roton determines $v_L$, the critical velocity is a decreasing function of density, a behavior unique to a DBEC.  Moreover, the simulations show a critical velocity that is somewhat smaller than $v_L$ at higher densities.  We attribute this to the role that the roton plays in the mechanical stability of a DBEC.

An ultracold, dilute DBEC containing $N$ atoms is well-modeled within mean-field theory by the time-dependent non-local Gross-Pitaevskii equation (GPE),
\begin{eqnarray}
\label{GPE}
i\hbar \frac{\partial \Psi(\rr,t)}{\partial t} = \left\{ -\frac{\hbar^2}{2M}\nabla^2 + U(\rr) + (N-1) \right. \nonumber \\ \left. \times \int d\rr^\prime V(\rr-\rr^\prime) |\Psi(\rr^\prime,t)|^2 \right\} \Psi(\rr,t)
\end{eqnarray}
where $\Psi(\rr,t)$ is the condensate wave function, normalized to unity; $\rr$ is the distance from the trap center; and $U(\rr) = \frac{1}{2}M\omega_\rho^2(\rho^2+\lambda^2 z^2)$ is the cylindrically symmetric harmonic trap potential with aspect ratio $\lambda = \omega_z / \omega_\rho$ where $\omega_z$ and $\omega_\rho$ are the axial and radial trap frequencies, respectively.  The two-body interaction potential for polarized dipoles with dipole moment $d$ and zero scattering length is~\cite{Yi00}
\begin{equation}
\label{int}
V(\rr-\rr^\prime) = 
%\frac{4\pi\hbar^2 a_s}{M} \delta(\rr-\rr^\prime) + 
d^2 \frac{1-3\cos^2{\theta}}{|\rr-\rr^\prime|^3},
\end{equation}
where $\theta$ is the angle between $\rr-\rr^\prime$ and the polarization axis.  We choose the polarization axis to be the trap axis, $\hat{z}$, so that the system is cylindrically symmetric.  To characterize the strength of the dipole-dipole interaction (ddi) in a DBEC, we define the dimensionless quantity $D = (N-1)\frac{M d^2}{\hbar^2 a_\rho}$ where $a_\rho = \sqrt{\hbar/M\omega_\rho}$ is the radial harmonic oscillator length of the trap.  The quantity $D$ then characterizes either the density of the gas or the dipole moments of the atoms in the gas.

We perturb this DBEC with a blue-detuned laser moving at constant velocity $v$, which amounts to adding a potential 
\begin{equation}
\label{laser}
U_\mathrm{las}(\rr,t) = \frac{U_0}{\sigma}\exp{\left[ \frac{-2(x^2+(y-y_\mathrm{ob}(t))^2)}{\sigma \tilde{w}_0^2} \right]}
\end{equation}
where $\sigma=1+(z/z_0)^2$, $z_0=\pi \tilde{w}_0^2/\lambda_\mathrm{las}$ is the Raleigh length, $\tilde{w}_0$ is the beam waist of the laser, $\lambda_\mathrm{las}$ is the wavelength of the laser, $y_\mathrm{ob}(t) = \Theta(t-t_0)[v(t-t_0)]$ describes the motion of the laser in the $y$-direction and $\Theta(t)$ is the Heaviside step function.  This potential describes a laser that is stationary until $t=t_0$, at which time it moves to the edge with velocity $\vec{v} = v\hat{y}$.

\begin{figure}
\includegraphics[width=\columnwidth]{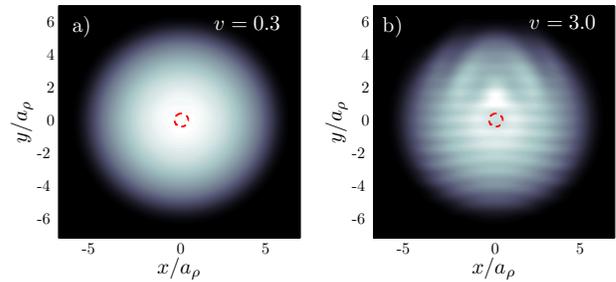}
\caption{\label{fig:condcomp} The density profiles of a DBEC with $D=124$ in a trap with aspect ratio $\lambda = 20$ after a blue-detuned laser with axis $\hat{z}$, beam waist $\tilde{w}_0 = 0.4 a_\rho$, $z_0 = 1.24 a_\rho$ and $U_0 = 2 \hbar \omega_\rho$ has traveled through the DBEC with velocity (a) $v=0.3$ $a_\rho \omega_\rho$ (b) $v=3.0$ $a_\rho \omega_\rho$. In (a), there are no visual excitations present in the system while in (b), excitations are clearly present, indicating the presence of a critical velocity for the system.  The $1/e^2$ contour of the laser is shown by the red dotted lines at the center of the condensates.}
\end{figure}

The effect of this blue-detuned laser on a DBEC is shown in Figure~\ref{fig:condcomp} for a DBEC with aspect ratio $\lambda = 20$, $D=124$, and a laser with $\tilde{w}_0 = 0.4 a_\rho$ and $U_0 = 2\hbar \omega_\rho$ where the chemical potential of the unperturbed condensate is $\mu = 26.3 \hbar \omega_\rho$.  We estimate the Landau critical velocity for this system to be $v_L\sim 1.5$ $a_\rho \omega_\rho$.  For a laser velocity less than this (Fig.~\ref{fig:condcomp}a), the condensate is completely unaffected whereas for a velocity larger than this (Fig.~\ref{fig:condcomp}b), quasiparticles are excited and the fluid would produce a net force on the moving laser.

To determine the Landau critical velocity $v_L$,  we calculate the condensate's quasiparticle spectrum by solving the Bogoliubov de Gennes (BdG) equations~\cite{Ronen06a}.  Due to cylindrical symmetry of the system, the condensate plus BdG quasiparticles can be written as
\begin{eqnarray}
\label{BdGansatz}
\Psi(\rr,t) \rightarrow \psi_0(\rho,z)e^{-i\mu t} + \sum_{j} \left\{ c_j(t)u_j(\rho,z)e^{i(m\varphi - \omega_j t)} \right. \nonumber \\ \left. + c^\star_j(t)v_j^\star(\rho,z)e^{-i(m\varphi-\omega_j t)} \right\} e^{-i\mu t}
\end{eqnarray}
where $\omega_j$ is the quasiparticle energy, $m$ is the projection of the quasiparticle momentum onto the $z$-axis and $\mu$ is the chemical potential of the ground state.  Here, $\psi_0(\rho,z)$ is the stationary condensate wave function, i.e.,  the solution of Eq.~(\ref{GPE}) with time-dependence $e^{-i\mu t}$, and is normalized to unity.  The coefficients $c_j(t)$ must be sufficiently small so that the BdG equations can be derived by linearizing the GPE about them.  Their time dependence describes slowly varying quasiparticle occupations (compared to $\omega_j^{-1}$) in out-of-equilibrium states.

In this formalism, the quasiparticles are characterized by their energies $\omega_j$ and $m$ quantum numbers.  However, in order to apply the Landau criterion to this system, the quasiparticles must be characterized by a  momentum, as well.  To do this, we calculate the expectation value of the  momentum, or $\langle k_\rho \rangle \equiv \sqrt{\langle k_\rho^2 \rangle}$, of the quasiparticles.  Using a Fourier-Hankel transform~\cite{Ronen06a}, we transform the modes into momentum-space and compute the expectation value of the linear momentum of the $j^\mathrm{th}$ quasiparticle in momentum-space representation,
\begin{equation}
\label{ky}
\langle k_\rho \rangle_j = \left\{\frac{\int d\kk \, k_\rho^2\left[ | \tilde{u}_j(\kk)|^2 + |\tilde{v}_j(\kk)|^2 \right]}{\int d\kk \left[ |\tilde{u}_j(\kk)|^2 + |\tilde{v}_j(\kk)|^2\right]} \right\}^\frac{1}{2},
\end{equation}
where we have time-averaged cross terms $\propto \cos{2\omega_j t}$ that oscillate on fast time scales~\cite{Morgan98}.  By associating these momenta to the excitation energies $\omega_j$, we determine a \emph{discrete} dispersion relation for this system.  The  Landau criteria for superfluid critical velocity is derived by applying conservation laws to translationally invariant fluids.  Since the fully trapped system that we consider here is translationally \emph{variant}, we apply the Landau criteria both to provide a hint as to where a critical velocity for quasiparticle excitations might be, and to test the application of this criterion to discrete systems.

\begin{figure}
\includegraphics[width=\columnwidth]{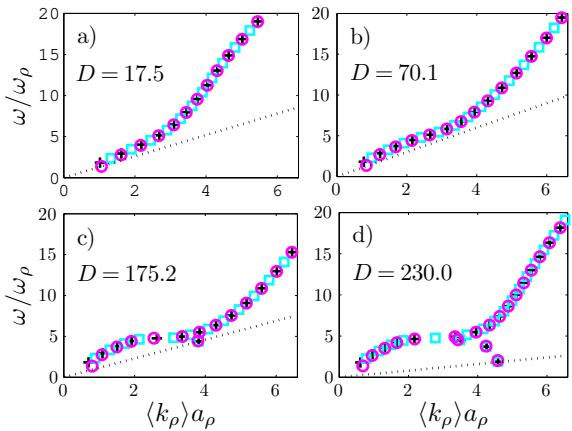}
\caption{\label{fig:discdisp} (Color online) The discrete BdG quasiparticle dispersions for a DBEC in a trap with aspect ratio $\lambda = 20$ for various values of $D$ showing $m=0$ (black $+$ sign), $m=1$ (teal squares) and $m=2$ (pink circles) quasiparticles.  As $D$ is increased, the dispersion develops a phonon-like character at low momenta and a roton-like character at intermediate momenta.  The slopes of the black dotted lines represent the corresponding Landau critical velocities for each $D$.}
\end{figure}

Figure~\ref{fig:discdisp} shows the discrete dispersion relations of a DBEC for various values of $D$.  For $D=0$ (not shown), the dispersion is given by the well known harmonic oscillator spectrum $\omega = n_\rho \omega_\rho$ with $\langle k_\rho \rangle = \sqrt{n_\rho+1} / a_\rho$ and $n_\rho = 0,1,2,..$.  However, as $D$ is increased, the spectrum changes to develop a phonon character at low-momenta and a roton character at intermediate momenta.  Indeed, for $D=175.2$, and more so for $D=230.0$, there are some quasiparticles that branch off from the dispersion towards lower energies and approach a momentum $\langle k_\rho \rangle \sim \sqrt{20} / a_\rho$, corresponding to the characteristic roton wavelength $\lambda_\mathrm{roton} \simeq 2\pi a_z$, where $a_z = \sqrt{\hbar / M\omega_z}$ is the axial harmonic oscillator length~\cite{Santos03a,Wilson08}.  The modes with similar momenta but larger energy, on the upper branch of the dispersion, exist in lower-density regions of the condensate while the quasiparticles on the roton branch exist in the high density center of the condensate.  Note that Figure~\ref{fig:discdisp} includes only quasiparticles with $m=0,1,2$.  

% The dispersions for quasiparticles with larger $m$, while sharing qualitative features with those of smaller $m$, shift towards larger momenta.  We attribute this to the fact that the presence of angular momentum in a fully trapped system changes the radial character of the quasiparticles and, thus, their $\langle k_\rho \rangle$.

In each case, the Landau critical velocity $v_L$ is determined according to Eq.~(\ref{landauvc}) as the slope of the shallowest line through the origin that intersects a point on the dispersion curve; these lines are indicated in the figure.  For smaller $D$, $v_L$ is determined by the low-momentum phonon-like modes where $\omega$ is linear in $\langle k_{\rho} \rangle$.  By contrast, for larger $D$, $v_L$ is determined by the low-lying roton mode and becomes a decreasing function of interaction strength in contrast to a BEC with only contact interactions, where $v_L$ grows as the square root of scattering length.

In evaluating $v_L$ from the discrete dispersion relation, we have ignored two excitations.  One is the unphysical $m=0$ Goldstone mode.  A second is the $m=1$ Kohn mode,  which has  eigenvalue $\omega_1 = \hbar \omega_\rho$ independent of interactions, and which corresponds to transverse sloshing of the condensate~\cite{Kohn61}.  The Kohn mode moves the condensate's center of mass rather than exciting quasiparticles relative to the center of mass, which would imply the breaking of superfluidity in a translationally invariant system.  We therefore ignore it here.  In any event, we find that the occupation of the Kohn mode is very small compared to the total condensate depletion.

We now compare $v_L$ as determined from the discrete dispersion relation with the onset of condensate depletion due to the laser having been moved through the DBEC.  To quantify the breaking of superfluidity in the simulations, we calculate the depletion of the condensate by finding the quasiparticle occupations, or the number of particles that are excited out of the condensed state.  In practice, this is achieved by calculating the amplitudes $c_j(t)$ in Eq.~(\ref{BdGansatz})~\cite{Ianeselli06} via the orthogonality relations of the BdG modes~\cite{Morgan98}, including their normalization $\int d\rr \left[ u^\star_{j}(\rr^\prime) u_{j^\prime}(\rr^\prime)- v^\star_{j}(\rr^\prime)v_{j^\prime}(\rr^\prime) \right] = \delta_{j j^\prime}$, to give
\begin{equation}
\label{coeff}
c_j(t) = \int d\rr^\prime \left[ u^\star_j(\rr^\prime)\Psi(\rr^\prime,t) - \Psi^\star(\rr^\prime,t) v_j^\star(\rr^\prime) \right] e^{i\omega_j t},
\end{equation}
where $\Psi(\rr,t)$ is the numerical solution of the time-dependent GPE with the blue-detuned laser potential.  The quasiparticle occupations are then given by $n_j(t) = |c_j(t)|^2\int d\rr^\prime(|u_j(\rr^\prime)|^2+|v_j(\rr^\prime)|^2)$.  In the simulations, the system evolves for a time $T$ after the laser has completely left the system.  We average the quasiparticle occupations for a time $T$ after this, giving the average excited state occupations $\bar{n}_j = \frac{1}{T} \int_0^T dt^\prime n_j(t^\prime)$.  We find that $T=5 \, \omega_\rho^{-1}$ is sufficient to converge these averages.

\begin{figure}
\includegraphics[width=\columnwidth]{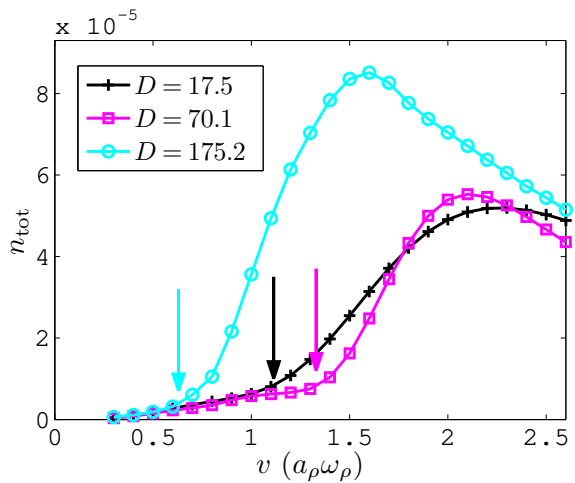}
\caption{\label{fig:occ} (Color online)  The occupations of the quasiparticles excited from a DBEC with aspect ratio $\lambda = 20$  by a blue-detuned laser moving with velocity $v$ (plotted on the horizontal axis) and with parameters $\tilde{w}_0 = 0.3a_\rho$, $U_0 = 0.4 \hbar \omega_\rho$ and $z_0 = 0.7 a_\rho$, for various values of $D$.  At a critical $v$ (indicated by the arrows), the occupations increase suddenly, indicating that the laser has excited quasiparticles in the system and superfluidity has been broken.}
\end{figure}

Figure~\ref{fig:occ} illustrates the total quasiparticle occupation $n_\mathrm{tot} = \sum_j \bar{n}_j$ as a function of laser velocity for various values of $D$ using the laser parameters $\tilde{w}_0 = 0.3 a_\rho$, $z_0=0.7 a_\rho$ and $U_0 = 0.4 \hbar \omega_\rho$.  For each $D$, $n_\mathrm{tot}$ stays very small until, at a certain critical velocity $v_\mathrm{crit}$, it begins to increase significantly.  Operationally, $v_\mathrm{crit}$ is determined by the intersection of linear fits below and above $v_\mathrm{crit}$.  Well above $v_\mathrm{crit}$, the occupations decrease with velocity since the laser spends proportionally less time in the system as its velocity is increased.

Notice that the overall depletion remains small with our weak laser.  We have deliberately remained in the perturbative limit with our simulations to uncover the basic physics without the complications of  large laser size.  Additionally, we have checked that these lasers are not sufficient to excite vortex states in the DBEC.  In practice, larger condensate depletion would be obtained from a repeated back-and-forth stirring, as was done in the MIT experiments, or from a wider, stronger laser.  While such a laser may spin off vortices in the condensate, thus defining a critical velocity smaller than $v_L$, the roton, for large enough $D$, would still determine the critical velocity.

\begin{figure}
\includegraphics[width=\columnwidth]{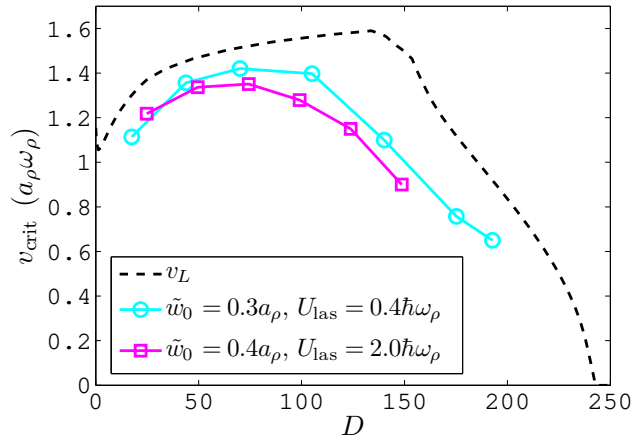}
\caption{\label{fig:vcrit} (Color online)  The superfluid critical velocities $v_\mathrm{crit}$ for dissipation due to the excitation of quasiparticles in a DBEC as a function of $D$.  The black dashed line represents the Landau critical velocity extracted from the discrete dispersion relations of the system.  The teal circles represent the results of numerical simulation for a laser with parameters $\tilde{w}_0 = 0.3 a_\rho$, $z_0 = 0.7 a_\rho$ and $U_0 = 0.4 \hbar \omega_\rho$ and the pink squares represent the results of numerical simulation for a laser with parameters $\tilde{w}_0 = 0.4 a_\rho$, $z_0 = 1.24 a_\rho$ and $U_0 = 2 \hbar \omega_\rho$.}
\end{figure}

Critical velocities determined from numerical simulations are presented in Figure~\ref{fig:vcrit} as a function of $D$. Results are shown for the comparatively weak ($U_0 = 0.4 \hbar \omega_\rho$) and strong ($U_0=2 \hbar \omega_\rho$) lasers.  Also shown for comparison is $v_L$ (dashed line) as determined from the discrete dispersion relations.  At small $D$, the critical velocity grows slightly as the phonon modes stiffen and the speed of sound increases.  This behavior is much like that of a BEC with purely contact interactions.

At higher density, the critical velocity instead decreases, due to the decreasing energy of the roton, and this is seen in both simulation and $v_L$.  The agreement is less perfect than in the phonon regime, however, with the simulated result coming in lower.  This is because the roton, being the collapse mechanism for DBECs in traps with larger aspect ratios, softens with increasing condensate density.  The presence of the laser in the DBEC serves to increase the density of the system, softening the roton and  thus decreasing the critical velocity of the condensate, just as a stationary laser leads a DBEC to instability~\cite{Asad09}.  For vanishingly small lasers, the critical velocities extracted from numerical simulation show increasingly better agreement with $v_L$.

Finally, it is worthwhile to consider measurements of critical velocities in experimentally accessible DBECs, such as the $^{52}\mathrm{Cr}$ system in Stuttgart~\cite{Griesmaier05a}. Consider $^{52}\mathrm{Cr}$ atoms whose scattering lengths have been tuned to zero in a trap with radial and axial frequencies $\omega_\rho = 2\pi\times 100$ Hz and $\omega_z = 2\pi\times 2000$ Hz, respectively.  This corresponds to a radial harmonic oscillator length of $a_\rho = 1.391\, \mu\mathrm{m}$, particle numbers  of $N \sim 570 D$ and  critical velocities in the range of 0.11 cm/s.  These circumstances suggest that it may be plausible to observe the decline of the superfluid velocity with $D$ for $N\gtrsim 8.5\times 10^4$ $^{52}\mathrm{Cr}$ atoms, and hence to exhibit directly the roton's influence on superfluidity.  This atom number corresponds to a maximum condensate density of $n_\mathrm{max} \simeq 9.5\times 10^{14}$ cm$^{-3}$, which, given the measured 3-body loss coefficient $L_3 = 2\times 10^{-28}$ cm$^6$/s~\cite{LahayeRev09}, should not produce significant losses over the time scales considered here.  Additionally, we have checked that, for sufficiently large $D$, the roton serves to determine $v_L$ for $^{52}\mathrm{Cr}$ DBECs with non-zero $s$-wave scattering lengths within the experimental uncertainty for $^{52}\mathrm{Cr}$, $-3 a_0 \leq a_s \leq 3 a_0$~\cite{Lahaye08PRL}, which is expected because these scattering lengths are sufficiently less than $^{52}\mathrm{Cr}$'s dipole length $a_{dd} \simeq 15 a_0$~\cite{LahayeRev09}.

The authors would like to acknowledge the financial support of the DOE and the NSF, and useful discussions with C. Raman.

\end{document}